%% file: paper.tex
\documentclass[acmsmall,screen,nonacm]{acmart}

\usepackage{microtype}
\usepackage{mathtools}
\usepackage{booktabs}
\usepackage{colortbl}
\usepackage{graphicx}
\usepackage{tabularx}
\usepackage{float}
\usepackage{algorithm}
\usepackage[noend]{algpseudocode}
\usepackage[nameinlink,noabbrev]{cleveref}

\makeatletter
\renewcommand{\theHALG@line}{\thealgorithm.\arabic{ALG@line}}
\makeatother

\theoremstyle{definition}

\newcommand{\tool}{NPTaylor}
\newcommand{\gsl}{GSL}
\newcommand{\autornp}{AutoRNP}

\setcopyright{none}
\acmYear{2027}
\acmDOI{}
\acmISBN{}
\acmConference[POPL 2027]{Proceedings of the ACM on Programming Languages, POPL 2027}{January 2027}{}
\acmBooktitle{}

\title[Taylor-Based Repair of Floating-Point Errors]{Taylor-Approximation-Based Repair of High Floating-Point Errors}

\author{Xin Yi}
\affiliation{%
  \institution{National University of Defense Technology}
  \city{Changsha}
  \country{China}}

\author{Liqian Chen}
\affiliation{%
  \institution{National University of Defense Technology}
  \city{Changsha}
  \country{China}}

\author{Hengbiao Yu}
\affiliation{%
  \institution{National University of Defense Technology}
  \city{Changsha}
  \country{China}}

\author{Shangwen Wang}
\affiliation{%
  \institution{National University of Defense Technology}
  \city{Changsha}
  \country{China}}

\author{Tao Tang}
\affiliation{%
  \institution{National University of Defense Technology}
  \city{Changsha}
  \country{China}}

\author{Chun Huang}
\affiliation{%
  \institution{National University of Defense Technology}
  \city{Changsha}
  \country{China}}

\author{Xiaoguang Mao}
\affiliation{%
  \institution{National University of Defense Technology}
  \city{Changsha}
  \country{China}}

\author{Ji Wang}
\affiliation{%
  \institution{National University of Defense Technology}
  \city{Changsha}
  \country{China}}
\renewcommand{\shortauthors}{Yi et al.}

\begin{abstract}
Mature numerical libraries can still lose many bits of accuracy on localized
ill-conditioned inputs, such as roots, poles, cancellation points, and branch
boundaries. Existing repair systems can localize such regions and synthesize
patches, but strict error thresholds often force many small pieces, long repair
times, or large generated code.

This paper presents \tool{}, a Taylor-approximation-based repair framework for
localized high floating-point errors in numerical C routines. Given a high-error
input and a localized repair domain, \tool{} searches over Taylor degree,
expansion center, segmentation, and floating-point evaluation strategy. It emits
guarded C patches that call the original routine outside the localized domain
and accepts a patch only after validating the emitted C code on the recorded
repair-domain samples, endpoints, and pre-repair peak input. The generated
Taylor evaluator uses coefficient splitting and compensated Horner evaluation to
reduce the evaluation error introduced by the patch itself.

On a 38-function one-input \gsl{} workload, both \tool{} and \autornp{} repair
all 114 configurations across three accuracy thresholds. \tool{} is
\textbf{3.33}$\times$ faster by geometric-mean offline repair time and emits
\textbf{0.162}$\times$ as many C bytes. In the strictest setting, the median
repair time is 3.74s for \tool{} and 81.97s for \autornp{}; the largest paired
repair-time speedup is \textbf{157.1}$\times$, and the largest strict patch is
47.5KB for \tool{} versus 5.90MB for \autornp{}. The repaired-code runtime
matrix gives \tool{} an overall geometric-mean speedup of
\textbf{1.56}$\times$ over the unpatched implementations, compared with
\textbf{1.35}$\times$ for \autornp{}. \tool{} also repairs localized
multi-input benchmark cases beyond \autornp{}'s implemented one-input workflow,
with 15/23 strict and 18/23 looser configurations repaired. These \gsl{} results
test the general repair method on library routines with known high-error
regions; they are local validation claims over the checked repair domains, not
full-domain proofs of the original routines.
\end{abstract}

\keywords{floating-point arithmetic, numerical programs, automated repair, GSL, Taylor approximation, compensated evaluation}

\begin{document}
\maketitle

\input{sections/01-introduction}
\input{sections/02-background}
\input{sections/03-motivating-examples}
\input{sections/03-problem-formulation}
\input{sections/04-taylibm-design}
\input{sections/05-evaluation}
\input{sections/06-related-work}
\input{sections/08-conclusion}
\input{sections/09-data-availability}

\bibliographystyle{ACM-Reference-Format}
\bibliography{references}

\end{document}

%% file: sections/01-introduction.tex
\section{Introduction}
\label{sec:introduction}

Numerical libraries are small pieces of code with unusually large
responsibility. Their functions sit beneath scientific computing, optimization,
simulation, sparse linear algebra, and machine-learning workloads, and their
numerical behavior can affect both accuracy and
reproducibility~\cite{goldberg1991every,sterbenz1974floating,Precimonious,lam2013automatically,sci1,sci2}.
The hardest failures are often not spread uniformly over the input space. A
mature routine may be accurate on most inputs but lose many bits near a root,
pole, cancellation point, or branch boundary. For users, these localized
failures are still library failures: the routine returns a representable value
far from the high-precision oracle-rounded result on inputs that the API
accepts.

Prior work has made high-error detection and repair increasingly automatic.
ATOMU~\cite{zoudetecting} detects high-error inputs using atomic conditions,
and FPCC~\cite{fpcc} uses chain conditions to find inputs that trigger large
floating-point error. \autornp{}~\cite{Yi:2019:EAR:3302515.3290369} repairs
high floating-point errors in numerical libraries by localizing suspicious
regions and searching for source-level patches. This is the closest baseline
for our local-repair setting, and its published artifacts support a matched
\gsl{} evaluation, but strict thresholds can make its repair search expensive
and its emitted patches large. In our strict repair experiment,
\autornp{} repairs \texttt{gsl\_sf\_bessel\_Y1} only after 9788.39s and emits
a 403KB patch for one localized interval; \tool{} repairs the same row in
139.23s with a 2.1KB patch.

This paper presents \tool{}, an automated framework for repairing localized
high floating-point errors in numerical C routines with generated Taylor
programs. The input to \tool{} is a high-error record, an original C routine, a
high-precision oracle policy, a localized repair domain, and an error threshold.
The output is a guarded C patch:
\[
  \textbf{if } x\in D \textbf{ then } E_D(x) \textbf{ else } P_0(x),
\]
where $P_0$ is the original implementation and $E_D$ is a generated Taylor
evaluator for the localized domain $D$. The repair preserves the original path
outside the guard and replaces only the high-error region. \tool{} accepts the
patch only after checking the emitted C program, including coefficient literals,
floating-point operation order, guard conditions, and fallback behavior.

Taylor approximation is useful here not because it is new, but because it
matches the locality of the repair problem. On a small interval or box, a
high-precision Taylor expansion around a selected center gives a compact
numerical form; the center, degree, and segmentation expose a small search
space; the same representation covers one-input intervals and multi-input
boxes; and fixed evaluation orders such as Horner and EFT--Horner map directly
to compact C. The generated expression is tied to the mathematical semantics of
the target function rather than only to a syntactic rewrite of the existing
implementation: on the selected domain, \tool{} approximates the high-precision
oracle and then validates the resulting floating-point program against the
repair threshold.

The key boundary is that \tool{} is a validation-driven repair system, not a
full-domain proof system. Its repaired patches are checked on
deterministic local validation sets that include sampled points, guard
endpoints, and the pre-repair peak input. A successful repair therefore means
that the emitted patch satisfies the stated threshold on the recorded
validation set for the localized domain; it does not prove continuity across
piece boundaries or a full-domain error bound for the original routine.

We evaluate this general repair method on baseline-aligned \gsl{} workloads.
On a 38-function one-input repair workload, both \tool{} and \autornp{} repair
all 114 configurations across three accuracy thresholds. \tool{} is
3.33$\times$ faster by geometric-mean repair time and emits 0.162$\times$ as
many C bytes. In the strictest setting, median repair time is 3.74s for
\tool{} versus 81.97s for \autornp{}, the largest paired repair-time speedup is
157.1$\times$, and the largest strict patch is 47.5KB for \tool{} versus
5.90MB for \autornp{}. The repaired-code runtime matrix also favors \tool{} in
aggregate: overall geometric-mean speedup over the unpatched implementations is
1.56$\times$ for \tool{} and 1.35$\times$ for \autornp{}. \tool{} also repairs
multi-input benchmark cases, a setting outside \autornp{}'s implemented
one-input workflow; it directly repairs up to 18/23 evaluated multi-input
cases.

\paragraph{Contributions}
This paper makes the following contributions.

\begin{itemize}
\item \textbf{A local repair formulation for generated C patches.} We formulate
high-error repair as domain-conditioned C-fragment generation with an explicit
checker contract. The formulation separates construction domains, validation
sets, guarded fallback behavior, and the local threshold claim made by a
reported repair.
\item \textbf{A Taylor-program repair strategy.} We design and implement
\tool{} as a checker-in-the-loop Taylor synthesis system that searches over
centers, degrees, segmentation, evaluator shapes, and coefficient splitting,
then makes its final decision on the emitted floating-point C code.
\item \textbf{A baseline-aligned evaluation on \gsl{} workloads.} We evaluate
\tool{} on \autornp{}-derived one-input repair configurations and a multi-input
\gsl{} capability study. \tool{} repairs all 114 matched one-input
configurations, reduces repair time and patch size relative to \autornp{},
improves aggregate runtime over the unpatched implementations, and repairs
localized multi-input cases beyond \autornp{}'s implemented workflow.
\end{itemize}

\paragraph{Paper Organization}
\Cref{sec:background} reviews the floating-point error metric, compensated
evaluation, and Taylor-approximation background used by the paper.
\Cref{sec:motivating-examples} gives a running repair example from \gsl{}.
\Cref{sec:problem} formalizes localized repair as checked C-fragment
generation. \Cref{sec:design} presents \tool{}'s repair search and C emission
pipeline. \Cref{sec:evaluation} describes the implementation, workloads, and
experimental results. \Cref{sec:related-work} positions \tool{} against repair,
approximation, and analysis systems, and \cref{sec:conclusion} concludes.

%% file: sections/02-background.tex
\section{Background}
\label{sec:background}

\subsection{Floating-Point Error for Repair}
\label{sec:background-fp}

Floating-point implementations approximate real-valued mathematical functions
with finite precision operations
\cite{goldberg1991every,sterbenz1974floating,kahan1996ieee,ieee7542019}. A
binary floating-point format $\mathbb{F}$ has precision $p$ and a bounded
exponent range. A finite normalized value has the form
\[
  x = (-1)^s m2^e,\qquad
  m = 1.b_1b_2\ldots b_{p-1},
\]
where $s$ is the sign bit and $e$ is an exponent in the format range.
Subnormal values use the minimum exponent with a leading significand bit of
zero. The format also contains signed zeros, infinities, and NaNs.

We use a high-precision oracle $f_{\mathrm{oracle}}$ as the mathematical
reference for a target function, following the common use of multiple-precision
libraries as reference implementations~\cite{mpfr}. Let
$\operatorname{round}_{\mathbb{F},\mathrm{RN}}$ denote round-to-nearest,
ties-to-even in the target format. For an implementation $P$, define
\[
  y_{\mathrm{ref}} =
  \operatorname{round}_{\mathbb{F},\mathrm{RN}}
  (f_{\mathrm{oracle}}(x)),
  \qquad
  y_{\mathrm{obs}} = P(x).
\]
Let $\operatorname{ord}_{\mathbb{F}}$ map non-NaN floating-point values to
monotonically increasing integers in numerical order. The repair error is the
distance between the observed and reference outputs in the ordered
floating-point value space:
\[
  \operatorname{ULPDist}(P,x)
  =
  \left|
  \operatorname{ord}_{\mathbb{F}}(y_{\mathrm{obs}})
  -
  \operatorname{ord}_{\mathbb{F}}(y_{\mathrm{ref}})
  \right|.
\]
The experiments report the logarithmic error metric
\[
  \operatorname{ErrBits}(P,x)=
  \begin{cases}
    0 & \operatorname{ULPDist}(P,x)=0,\\
    \log_2(\operatorname{ULPDist}(P,x)) &
      \operatorname{ULPDist}(P,x)>0.
  \end{cases}
\]
Larger values indicate a larger representable-value distance from the
oracle-rounded reference. The repair workload in \cref{sec:evaluation-setup}
selects high-error inputs with $\operatorname{ErrBits}(P_0,x)>32$.

\subsection{Localized Repair}
\label{sec:background-repair}

Localized floating-point repair has three roles. A detector finds high-error
inputs; a localizer expands such an input into a repair domain where the
problem recurs; and a repair backend synthesizes code for that domain.
\autornp{} uses perturbation-based testing (PTB) to localize intervals and
then emits source-level patches for the localized region
\cite{Yi:2019:EAR:3302515.3290369}. \tool{} reuses this localization interface
for a baseline-aligned comparison: both systems repair the same localized
domains and are compared on the generated backend time, patch size, post-repair
sampled error, and patched-code runtime.

A guarded repair has the form
\[
  P_{\mathrm{patch}}(x)=
  \begin{cases}
    E_D(x) & x\in D,\\
    P_0(x) & x\notin D,
  \end{cases}
\]
where $D$ is the localized repair domain, $E_D$ is the generated evaluator, and
$P_0$ is the original implementation. This form localizes behavioral change to
the suspicious region. It also makes the validation boundary explicit:
post-repair claims are about the checked inputs inside the guard, while the
outside path is the original routine plus guard overhead.

\subsection{Error-Free Transformations and Compensated Evaluation}
\label{sec:background-eft}

Error-free transformations (EFTs) expose the rounding error of a floating-point
operation as an explicit floating-point residual
\cite{Dekker:1971:FTE:2716631.2717032,accdotpro}. For addition, \textsc{TwoSum}
returns a rounded sum and a compensation term:
\[
  (s,e_s)=\textsc{TwoSum}(a,b), \qquad
  s=\operatorname{fl}(a+b), \qquad s+e_s=a+b,
\]
assuming the operation does not overflow and the standard conditions for the
transformation hold. For multiplication, \textsc{TwoProd} similarly computes
\[
  (p,e_p)=\textsc{TwoProd}(a,b), \qquad
  p=\operatorname{fl}(ab), \qquad p+e_p=ab.
\]
Dekker's construction obtains the product residual by splitting significands
into high and low parts, while hardware fused multiply-add can compute the
residual directly as
\[
  e_p=\operatorname{fma}(a,b,-p)
\]
when FMA is available~\cite{fmaIEEE}.

EFTs are useful for generated polynomial code because approximation error and
floating-point evaluation error interact. A high-precision Taylor coefficient
$c$ can be stored as a pair
\[
  c^h=\operatorname{fl}(c), \qquad
  c^\ell=\operatorname{fl}(c-c^h),
\]
so the emitted evaluator can use the high part in the main computation and feed
the low part into a compensation path. In an EFT--Horner evaluator, each Horner
multiply-add step records multiplication and addition residuals with
\textsc{TwoProd} and \textsc{TwoSum}; the final value combines the main Horner
result with accumulated residual terms. This improves the numerical behavior of
the emitted Taylor evaluator, but it is still only a candidate construction:
\tool{} accepts a repair only after checking the emitted C patch.

\subsection{Taylor Approximation as Local Repair Code}
\label{sec:background-taylor}

Taylor polynomials are a classical way to approximate smooth functions on a
small neighborhood around an expansion point, alongside other standard
polynomial-approximation tools used in numerical software
\cite{Abramowitz:1974:HMF:1098650,Knuth:1997:ACP:270146,hart1968computer,muller2016elementary}.
For a univariate function $f$ and center $a$, the degree-$d$ template is
\[
  T_{a,d}(x) = \sum_{i=0}^{d} c_i (x-a)^i.
\]
For a multi-input function $f:\mathbb{R}^n\to\mathbb{R}$, the corresponding
template uses multi-indices. Let $\boldsymbol{x}=(x_1,\ldots,x_n)$,
$\boldsymbol{a}=(a_1,\ldots,a_n)$, and
$\boldsymbol{\alpha}=(\alpha_1,\ldots,\alpha_n)\in\mathbb{N}^n$, with
$|\boldsymbol{\alpha}|=\alpha_1+\cdots+\alpha_n$,
$\boldsymbol{\alpha}!=\alpha_1!\cdots\alpha_n!$, and
$(\boldsymbol{x}-\boldsymbol{a})^{\boldsymbol{\alpha}}
=\prod_{j=1}^n (x_j-a_j)^{\alpha_j}$. The total-degree-$d$ Taylor polynomial is
\[
  T_{\boldsymbol{a},d}(\boldsymbol{x})
  =
  \sum_{|\boldsymbol{\alpha}|\le d}
  C_{\boldsymbol{\alpha}}
  (\boldsymbol{x}-\boldsymbol{a})^{\boldsymbol{\alpha}},
  \qquad
  C_{\boldsymbol{\alpha}}
  =
  \frac{1}{\boldsymbol{\alpha}!}
  \frac{\partial^{|\boldsymbol{\alpha}|}f(\boldsymbol{a})}
       {\partial x_1^{\alpha_1}\cdots\partial x_n^{\alpha_n}} .
\]
Thus a localized box for a multi-input routine can be represented by the same
kind of local polynomial object as a one-input interval, only with
multi-indexed coefficients.

The same real polynomial can produce different target results when evaluated
with different operation orders, intermediate precisions, or compensated
arithmetic schemes. \tool{} therefore treats the evaluation scheme as part of
the candidate. Horner evaluation provides a compact fixed order
\cite{cajori1911horner,mulhorner}. EFT--Horner combines the Taylor
representation with the residual terms above: coefficients are split into high
and low parts, each multiplication and addition exposes a residual, and the
residual path is accumulated into the final value. The emitted program includes
the chosen center, coefficients, degree, guard, and operation order.

%% file: sections/03-motivating-examples.tex
\section{Motivating Example}
\label{sec:motivating-examples}

\subsection{Localized Repair: \texttt{gsl\_sf\_bessel\_Y1}}
\label{sec:motivating-repair}

The running example is the order-one Bessel function of the second kind,
$Y_1(x)$, which is the solution of
\[
  x^2 y'' + x y' + (x^2-1)y = 0
\]
selected by the Bessel-$Y$ boundary condition. For the $H_\epsilon$ row,
\autornp{}'s PTB stage localizes the high-error interval
$D=[30.6047,30.6318]$ with peak input $x=30.6183$ near its center. The row
threshold is $H_\epsilon=5.13$ ErrBits.

\tool{} receives the same local domain and applies the degree-first,
then-split repair policy of \cref{sec:design-repair}. It first increases the
Taylor degree for the current interval; if no candidate passes the error
threshold, it subdivides the interval, reanchors the failed subdomains, and
repeats this process until the checker finds no validation input whose error
exceeds $H_\epsilon$. In this case, the current artifact row accepts a
one-piece degree-7 Taylor polynomial evaluated with EFT--Horner and finishes
in 139.23s.

Concretely, the repair instance is the original \texttt{bessel\_Y1} C routine,
the interval $D$, the peak anchor, the threshold $H_\epsilon$, and a checker
input set containing 10,000 deterministic samples plus endpoints and the peak.
\textsc{TryDegrees} emits the guarded degree-$d$ C candidate for each
$d=0,\ldots,7$ and checks that concrete wrapper. A failure returns the sampled
inputs above $H_\epsilon$; only then does the policy split the corresponding
interval. The accepted row needs no split, but the state passed between
iterations is explicit rather than an abstract ``increase degree'' instruction.

\begin{figure}[H]
\centering
\includegraphics[width=0.82\textwidth]{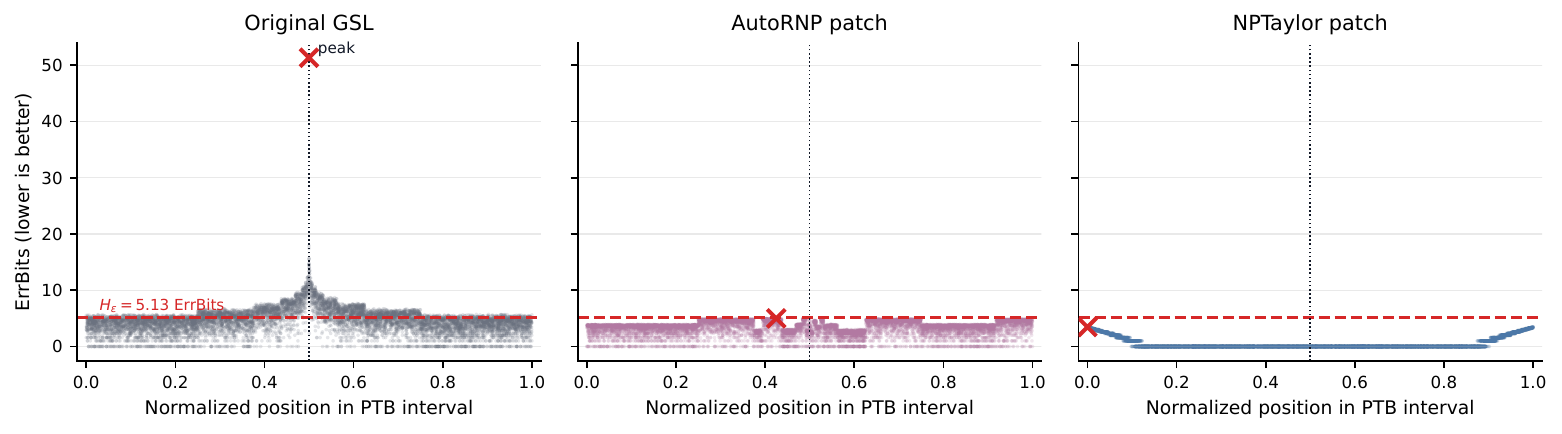}
\caption{\texttt{gsl\_sf\_bessel\_Y1} repair at the $H_\epsilon=5.13$ ErrBits threshold. All panels use the same PTB-localized interval and
10,000 fixed-seed samples. The original routine peaks at 51.31 ErrBits, while the repaired samples stay below
the bound: \autornp{} peaks at 5.00 ErrBits and \tool{} peaks at 3.46 ErrBits.}
\Description{A three-panel scatter plot over normalized position in the PTB
interval. A red dashed line marks H epsilon equals 5.13 ErrBits, and each panel
marks its sampled maximum with a red cross. The original-routine panel has many
points above the bound and a peak of 51.31 ErrBits near the interval center.
The AutoRNP patch peaks at 5.00 ErrBits and the NPTaylor patch peaks at 3.46
ErrBits, both below the bound.}
\label{fig:motivating-repair-y1}
\end{figure}

\Cref{fig:motivating-repair-y1} illustrates the behavior behind this policy.
A Taylor polynomial is most accurate near its expansion center, so the repaired
error is close to zero near the middle of the interval and increases toward its
ends. Had either end exceeded the threshold, \tool{} would have split that
region and repaired the resulting subintervals separately. The original routine
peaks at 51.31 ErrBits, whereas the sampled \tool{} patch peaks at
3.46 ErrBits. \autornp{} follows a related refine-until-accepted workflow, but
uses fixed-degree piecewise quadratic approximations rather than raising the
approximation degree. Its matched repair takes 9788.39s (70.3$\times$ the
\tool{} backend time) and emits a 403.3KB patch with 2,542 tabulated grid
points; this is about 97$\times$ the 4,177-byte original
\texttt{bessel\_Y1.c} source. In contrast, the \tool{} patch is only 2,131
bytes.

Over the high-error interval
$D=[30.6047,30.6318]$, the original path evaluates the amplitude--phase
form
\[
  Y_1(x) \approx -\frac{0.75+C_a(z)}{\sqrt{x}}
  \cos\!\left(x-\frac{\pi}{4}+\frac{C_\theta(z)}{x}\right),
  \qquad z=\frac{32}{x^2}-1,
\]
where $C_a$ and $C_\theta$ are Chebyshev correction series with propagated
error bounds. The guarded \tool{} path instead evaluates one degree-7 Taylor
polynomial $P_7(x-x_{\mathrm{peak}})$, making it 4.93$\times$ faster than the
original inside the interval, while \autornp{} reaches 1.15$\times$ on the
same stream. The \tool{} mixed-input speedup is 1.63$\times$ despite a 2.6\%
guard overhead outside the interval.

%% file: sections/03-problem-formulation.tex
\section{Problem Formulation}
\label{sec:problem}

This section states the repair interface that \tool{} must satisfy; the search
procedure appears in \cref{sec:design}. \tool{} does not infer repairs from
arbitrary C code in isolation. A concrete repair instance is supplied by a
detector/localizer front end and contains the original routine, localized
domain, oracle policy, threshold, and validation policy.

\subsection{Domain-Conditioned Repair Interface}
\label{sec:problem-domain-conditioned}

Let $P_0$ be the original floating-point implementation and
$f_{\mathrm{oracle}}$ be the high-precision mathematical reference. A repair
instance supplies a target floating-point format $\mathbb{F}$, one or more
localized construction domains $\mathcal{D}_{\mathrm{loc}}$, a checked
validation set $D_{\mathrm{val}}$, target C floating-point semantics $S_C$, and
an error threshold $\epsilon$. The construction domains come from the repair
front end, such as PTB localization around a high-error input; the validation
set is the finite set on which the final emitted patch is checked.

\tool{} emits local evaluators $E_1,\ldots,E_m$ and composes them with ordered
guards and the original fallback:
\[
  P_{\mathrm{patch}} = C_{\mathrm{repair}}[E_1,\ldots,E_m].
\]
For one domain $D$, this context is
\[
  C_{\mathrm{repair}}[E](x)=
  \textbf{if }g_D(x)\textbf{ then }E(x)\textbf{ else }P_0(x),
\]
where $g_D$ is the emitted guard for the localized domain. With multiple
pieces, the guards are ordered and the fallback remains $P_0$. The checked
object is the generated C program, not only the mathematical Taylor polynomial
used to construct it. The common checker contract is:
\[
  \mathsf{Check}(P_{\mathrm{patch}},\Phi_{\mathrm{repair}},
  D_{\mathrm{val}},f_{\mathrm{oracle}},S_C)=\mathsf{pass}.
\]

\subsection{Repair Acceptance Condition}
\label{sec:problem-acceptance}

Using the ErrBits metric from \cref{sec:background-fp}, the evaluated repair
condition is
\[
\Phi_{\mathrm{repair}}(P,x,f_{\mathrm{oracle}})
\equiv
\operatorname{ErrBits}(P,x) \le \epsilon.
\]
The checker applies this predicate to its finite recorded validation set. In
our experiments, this set includes deterministic repair-domain samples, guard
endpoints, and the pre-repair peak input identified before synthesis. A repair
is reported only if the final emitted patch, including all guards and fallback
paths, satisfies the threshold on this set.

The construction domain and checked domain play different roles.
$\mathcal{D}_{\mathrm{loc}}$ guides Taylor construction, guard generation, and
subdivision; $D_{\mathrm{val}}$ is the evidence set for the reported result.
The final claim is validation-relative: when \tool{} reports success, it
reports that the emitted patch passed the checker contract for the stated
localized domain, validation set, oracle policy, threshold, and C semantics.

\subsection{Claim Boundary}
\label{sec:problem-claim-boundary}

\tool{}'s repair claim is local. It does not establish a proof over every
representable input in the original routine, and it does not prove
continuity between adjacent real-polynomial pieces. At a boundary, the emitted
ordered guard selects one C path and the selected output is checked on the
validation inputs that include guard endpoints. The repair experiments check the
scalar returned value of the generated patch wrapper; they do not claim
preservation of every auxiliary observable such as status codes, \texttt{errno},
or floating-point exception flags. These limits are intentional: they keep the
paper's evidence aligned with the checker actually used to accept patches.

%% file: sections/04-taylibm-design.tex
\section{\tool{} Design}
\label{sec:design}

\subsection{Overview}
\label{sec:design-overview}

\Cref{sec:problem} defines the checker contract that \tool{} must satisfy.
\tool{} searches for an executable Taylor program that satisfies this contract
after being lowered to complete C. The pipeline has five stages:

\begin{enumerate}
\item \textbf{Localize the high-error region.} The front end receives a
high-error record and uses PTB localization to construct an interval or box
around the reported peak input.
\item \textbf{Build a Taylor candidate.} The builder fixes the expansion
center, degree, and pieces selected by the policy and computes high-precision
Taylor coefficients for the oracle target.
\item \textbf{Fix the floating-point evaluator.} The generator selects the
evaluator shape, coefficient splitting, C literals, and operation order,
thereby turning the mathematical approximation into a concrete floating-point
program.
\item \textbf{Emit guarded C.} The evaluator is composed with guards and the
fallback call to the original routine.
\item \textbf{Check the emitted patch.} The checker executes the exact C
program that \tool{} may return, rather than checking only its real-number
polynomial.
\end{enumerate}

A passing check produces a guarded repair patch. A failure returns sampled
inputs whose errors still exceed the threshold, and the search policy refines
the Taylor descriptor. Consequently, every refinement is validated after
lowering to complete C.

The loop communicates through a compact \emph{Taylor-program descriptor}:
\[
  q = \langle B, a, d, \sigma, g_B\rangle .
\]
Here $B$ is the current construction region, $a$ is the expansion center, $d$
is the Taylor degree, $\sigma$ is the fixed floating-point evaluator shape, and
$g_B$ is the emitted guard. The descriptor is not a separate specification
language; it is the interface between three roles. The repair policy selects
and refines $q$, the builder lowers $q$ to a complete C patch, and the checker
alone decides whether the patch is accepted.

\subsection{Repair Policy}
\label{sec:design-repair}

The repair policy keeps two domains separate: the PTB-localized construction
region ($D_{\mathrm{loc}}$) is used to construct guards and Taylor pieces,
whereas the finite validation set ($D_{\mathrm{val}}$) is used to check the
emitted patch. The policy tries degrees before splitting a failed region and
validates the merged guarded patch at the end. In the
\texttt{gsl\_sf\_bessel\_Y1} running example, the initial descriptor has the
form
\[
q_{\mathrm{Y1}} =
\langle D_{\mathrm{loc}},x_{\mathrm{peak}},d,
\mathrm{EFT}\text{-}\mathrm{Horner},g_D\rangle ,
\]
where $D_{\mathrm{loc}}$ is the PTB interval, $x_{\mathrm{peak}}$ is the
reported high-error input used as the first anchor, and $g_D$ is the repair
guard.

In \cref{alg:repair-refinement}, the worklist ($W$) holds region--anchor pairs,
the current pair is a region $B$ and anchor $a$, and $g_B$ is $B$'s guard. The
algorithm accumulates accepted local descriptors in $\mathcal{Q}$, uses
$D_B^{\mathrm{val}}$ for the current region's validation inputs, and records a
candidate local descriptor and its checker result as $(q_B,v)$.

\begin{algorithm}[H]
\small
\caption{Repair search policy: try degree, then split failed regions}
\label{alg:repair-refinement}
\begin{algorithmic}[1]
\Require high-error record $r$, original program $P_0$, oracle policy $O$,
repair condition $\Phi_{\mathrm{repair}}$, validation-set policy $V$,
C semantics $S_C$, degree bound $d_{\max}$, piece bound $m_{\max}$
\Ensure guarded patch $P_{\mathrm{patch}}$ or failure
\Statex \textbf{Phase 1: Localize the region to repair.}
\State $x_{\mathrm{peak}} \gets \Call{PeakInput}{r}$
\State $D_{\mathrm{loc}} \gets
  \Call{PTBLocalize}{P_0,O,x_{\mathrm{peak}}}$
\State $W \gets [(D_{\mathrm{loc}},x_{\mathrm{peak}})]$;
       $\mathcal{Q} \gets \emptyset$; $D_{\mathrm{val}}\gets\emptyset$
\Statex \textbf{Phase 2: Cover each region with a checked Taylor piece.}
\While{$W \ne \emptyset$}
  \State $(B,a) \gets \Call{Pop}{W}$
  \State $g_B \gets \Call{Guard}{B}$
  \State $D_B^{\mathrm{val}} \gets \Call{ValidationInputs}{V,B,g_B,r}$
  \State $(q_B,v) \gets \Call{TryDegrees}{O,P_0,B,a,g_B,
    D_B^{\mathrm{val}},\Phi_{\mathrm{repair}},S_C,d_{\max}}$
  \If{\Call{RepairAccept}{$v$}}
    \State $\mathcal{Q} \gets \mathcal{Q}\cup\{q_B\}$
    \State $D_{\mathrm{val}}\gets D_{\mathrm{val}}\cup D_B^{\mathrm{val}}$
  \ElsIf{$|\mathcal{Q}|+|W|+1\ge m_{\max}$}
    \State \Return \Call{Failure}{$B$}
  \Else
    \State $W \gets W \cup \Call{SubdivideAndAnchor}{B,a,v}$
  \EndIf
\EndWhile
\Statex \textbf{Phase 3: Merge pieces and validate the final patch.}
\State $P_{\mathrm{patch}} \gets
  \Call{ComposeRepair}{P_0,\mathcal{Q}}$
\State $v_{\mathrm{final}}\gets \Call{Check}{P_{\mathrm{patch}},
  \Phi_{\mathrm{repair}},D_{\mathrm{val}},O,S_C}$
\State \Return \Call{AcceptOrFail}{$P_{\mathrm{patch}},v_{\mathrm{final}}$}
\end{algorithmic}
\end{algorithm}

\textsc{TryDegrees} repeatedly builds and checks, for $d=0,\ldots,d_{\max}$,
candidate local descriptors of the form
\[
\langle B,a,d,\mathrm{EFT}\text{-}\mathrm{Horner},g_B\rangle
\]
using the temporary context
$C_B[e](x)=\textbf{if }g_B(x)\textbf{ then }e(x)\textbf{ else }P_0(x)$.
It returns the first accepted local descriptor or the last checker result $v$.
Only after this sweep fails does the policy split the region and choose new
anchors from checker-reported difficult subdomains. In one-input repair this
split is trisection; the implementation also supports multi-input strip and
thirds-based box splitting. The evaluated configuration uses $d_{\max}=7$ and
piece bounds of 81 for one-input and 1000 for multi-input repair.

\subsection{Taylor Candidate Construction}
\label{sec:design-construction}

For each descriptor, \tool{} evaluates the high-precision oracle near the
selected anchor and derives Taylor coefficients for the local target. The
center is initially the detector-reported peak input because the peak is the
most informative point in the localized region. After subdivision, the policy
can reanchor a failed subdomain at the checker-reported difficult input or at
the subdomain center.

The emitted evaluator uses coefficient splitting and EFT--Horner. Each
coefficient is rounded to a high part and a low residual. The high parts form
the main Horner chain, while the residual terms from \textsc{TwoProd},
\textsc{TwoSum}, and low coefficients are accumulated and added back. The goal
is not to prove a symbolic error bound for the Taylor expression; it is to
produce a candidate whose actual floating-point execution has enough numerical
headroom to pass the local checker.

\subsection{C Emission and Validation}
\label{sec:design-emission}

After one or more local descriptors are accepted, \tool{} emits a merged patch
with ordered guards. Each guard selects the Taylor evaluator for its localized
piece; inputs outside all guards fall back to the original routine. The final
validation rechecks the merged patch rather than assuming that local
acceptance remains valid after composition. This catches mistakes in guard
ordering, coefficient literal emission, and fallback integration.

Changing compiler flags, target ISA, floating-point environment, or validation
policy can change the behavior of floating-point C. For that reason, \tool{}'s
acceptance decision is tied to the emitted source and the stated checking
environment. A repair result is reported only after that concrete source passes
the local validation contract.

%% file: sections/05-evaluation.tex
\section{Implementation and Evaluation}
\label{sec:evaluation}

\definecolor{rqanswerbg}{RGB}{244,248,252}
\definecolor{rqanswerframe}{RGB}{142,164,190}
\newcommand{\rqanswer}[2]{%
\par
\noindent\begingroup
\setlength{\fboxsep}{7pt}%
\setlength{\fboxrule}{0.5pt}%
\fcolorbox{rqanswerframe}{rqanswerbg}{%
\begin{minipage}{\dimexpr\linewidth-2\fboxsep-2\fboxrule\relax}
\textbf{#1.} #2
\end{minipage}}%
\endgroup\par}
\DeclareRobustCommand{\metric}[1]{\textbf{#1}}
\DeclareRobustCommand{\Leps}{\ensuremath{L_\epsilon}}
\DeclareRobustCommand{\Meps}{\ensuremath{M_\epsilon}}
\DeclareRobustCommand{\Heps}{\ensuremath{H_\epsilon}}

\subsection{Implementation}
\label{sec:implementation}

We implemented \tool{} as a prototype in Python, C, and C++. The Python
orchestrator manages detection, PTB localization, search, candidate selection,
and artifact generation; performance-critical oracle and validation paths use
C/C++ with MPFR/GMP support~\cite{mpfr}. The generator emits ordinary C
patches with domain guards and fallback calls to the original routines.
All validation and benchmarking use the emitted C source, not an idealized
symbolic approximation.

The implementation follows the workflow in \cref{sec:design}. The detector
consumes high-error inputs reported by FPCC and uses \autornp{}'s PTB algorithm
to localize a one-dimensional interval or a low-dimensional box around the
reported peak. The deriver builds Taylor approximations over those domains and
fixes the expansion center, degree, segmentation, and evaluation order. Repair
candidates use coefficient splitting and EFT--Horner. The generator then emits
a guarded patch and invokes the local checker on the concrete compiled wrapper.

\subsection{Evaluation Overview}
\label{sec:evaluation-overview}

The evaluation is organized around two questions:

\begin{description}
\item[RQ1] \textit{Repair effectiveness.} Can \tool{} repair localized
high-error cases on \gsl{} benchmark programs?
\item[RQ2] \textit{Repair quality and cost.} How large are the generated
patches, how long does repair take, and what runtime impact do the patches have?
\end{description}

One-input repair is paired with \autornp{} on shared PTB domains. Multi-input
repair is a \tool{}-only capability study because \autornp{} does not support
multi-input repair in its implemented workflow.

\subsection{Experimental Setup}
\label{sec:evaluation-setup}

\paragraph{Benchmarks.}
FPCC screens 167 \gsl{} special-functions~\cite{gslmanual} and retains ErrBits
$>32$, yielding 38 one-input and 23 multi-input records (17 two-, 2 three-, and
4 four-input). We choose \gsl{} because it is a widely used
scientific-computing library, so its routines provide practically
representative numerical workloads. Prior floating-point error-detection and
repair studies also use \gsl{} programs, enabling comparison with established
benchmarks and results
\cite{conf/icse/ZouWXZSM15,yi2017automated,Yi:2019:EAR:3302515.3290369}.

\paragraph{Repair thresholds.}
We reuse \autornp{}'s threshold rule
$q_\alpha=b_{\mathrm{avg}}+\alpha(b_{\max}-b_{\mathrm{avg}})$, where
$b_{\mathrm{avg}}=\max(\log_2 A,0)$ for mean ULP error $A$ and
$\alpha=0.1,0.2,0.3$ gives \Heps{}, \Meps{}, and \Leps{} from strict to
loose; larger $\alpha$ relaxes the allowed error.

\paragraph{Checking criteria.}
A repair is counted as successful when the backend reports a completed patch
under the shared PTB threshold (\texttt{repaired} for \tool{} and \texttt{ok}
for \autornp{}). Separately, the reported post-repair mean and maximum ErrBits
are computed on a deterministic 10,000-point repair-domain sample. The set
includes guard-boundary inputs and the error-maximizing input identified before
repair. Thus, the reported error metrics are local diagnostics over the emitted
guard rather than a full-domain proof of the surrounding original routine.

\paragraph{Metrics.}
Let $\mathcal{V}$ be the validation set and let $e(x)$ be the ULP error of the
repaired program at $x$. \textbf{Maximum repaired error} is
$E_{\max}=\max_{x\in\mathcal{V}}\log_2 \max(e(x),1)$ (ErrBits), which captures
the worst sampled post-repair error. \textbf{Mean repaired error} is
$E_{\mathrm{avg}}=|\mathcal{V}|^{-1}\sum_{x\in\mathcal{V}}\log_2 \max(e(x),1)$,
which summarizes typical post-repair ErrBits over the samples.
\textbf{Runtime speedup} is $S_{\mathrm{run}}=T_{\mathrm{orig}}/T_{\mathrm{rep}}$,
where larger values mean faster repaired code. \textbf{Repair time} is
$T_{\mathrm{repair}}$, the elapsed synthesis-and-emission time excluding the
shared PTB stage, where smaller is better. \textbf{Patch size} is
$B_{\mathrm{patch}}$, the emitted C-code size in bytes, where smaller patches
are preferable. \tool{} searches degrees 0--7 with at most 81 one-input or
1000 multi-input pieces. Runtime uses 2,000,000 timed calls after 50,000
warmups, with three repeats on inside, outside, and mixed input streams; we
aggregate speedups by geometric mean.

RQ1--RQ2 use an Intel Core i9-13900 host with 24 cores, 125 GiB RAM, Ubuntu
22.04/Linux 6.5, and GCC 11.4. The multi-input study reports PTB localization,
direct repair, and sensitivity to a cap on the number of input tuples in a
repair box. The cap controls synthesis tractability, not the error threshold.
Validation and timing use the same wrapper: dispatch to the Taylor patch inside
the box and otherwise call the original routine.

\subsection{RQ1: Repair Effectiveness}
\label{sec:evaluation-rq1}

\begin{figure}[H]
\centering
\includegraphics[width=\textwidth]{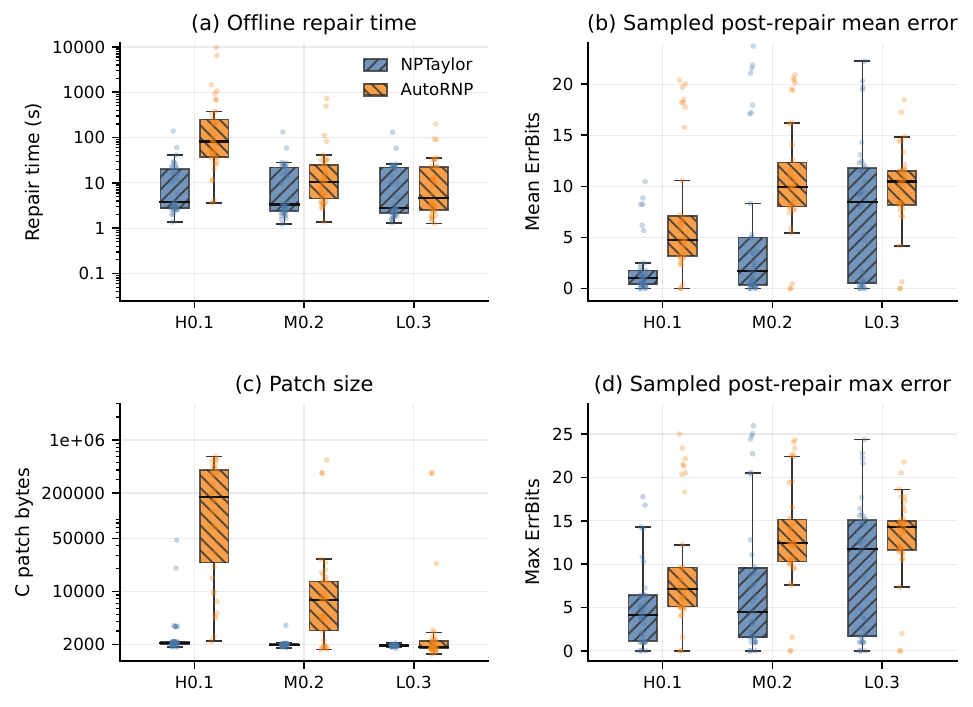}
\caption{One-input repair distributions over 38 shared mixed-PTB functions at
three thresholds. Panels show offline repair time, sampled post-repair mean
ErrBits, C patch bytes, and sampled post-repair max ErrBits. Lower is better in
all panels. Strict repair is where \tool{} most clearly reduces time and patch
size.}
\Description{Four boxplots showing repair time, sampled post-repair mean
error, C patch bytes, and sampled post-repair maximum error for NPTaylor and
AutoRNP.}
\label{fig:eval-repair-1d-summary}
\end{figure}

On the shared mixed-PTB workload, both \tool{} and \autornp{} repair all 38
standard one-input functions at all three thresholds. The separately excluded
\texttt{gsl\_sf\_psi\_1} pole case exposes a representation boundary analyzed
below. \Cref{fig:eval-repair-1d-summary} and
\cref{tab:repair-level-summary} show where the successful repairs become
expensive, while the multi-input results isolate localization from synthesis
failure.

\begin{table}[H]
\centering
\caption{Direct one-input repair comparison over 38 functions on shared mixed
PTB. Repair time excludes shared PTB; lower time, error, and bytes are better.
The last column reports the \tool{} advantage, computed as
\autornp{}/\tool{} for lower-is-better metrics.}
\label{tab:repair-level-summary}
\scriptsize
\setlength{\tabcolsep}{4.0pt}
\begin{tabular}{llrrl}
\toprule
\metric{Threshold} & \metric{Metric} & \tool{} & \autornp{} &
\metric{\tool{} advantage} \\
\midrule
\,\Heps{} & Pass & \textbf{38/38} & \textbf{38/38} & tie \\
          & Repair time & \textbf{3.74s} & 81.97s & \textbf{21.9$\times$ faster} \\
          & C patch bytes & \textbf{2.1K} & 177.4K & \textbf{84.5$\times$ smaller} \\
          & Avg post ErrBits & \textbf{1.00} & 4.78 & \textbf{4.8$\times$ lower} \\
          & Max post ErrBits & \textbf{4.16} & 7.18 & \textbf{1.7$\times$ lower} \\
\midrule
\,\Meps{} & Pass & \textbf{38/38} & \textbf{38/38} & tie \\
          & Repair time & \textbf{3.32s} & 10.40s & \textbf{3.1$\times$ faster} \\
          & C patch bytes & \textbf{2.0K} & 7.7K & \textbf{3.9$\times$ smaller} \\
          & Avg post ErrBits & \textbf{1.70} & 9.96 & \textbf{5.9$\times$ lower} \\
          & Max post ErrBits & \textbf{4.52} & 12.42 & \textbf{2.7$\times$ lower} \\
\midrule
\,\Leps{} & Pass & \textbf{38/38} & \textbf{38/38} & tie \\
          & Repair time & \textbf{2.78s} & 4.61s & \textbf{1.7$\times$ faster} \\
          & C patch bytes & \textbf{1.9K} & \textbf{1.9K} & 1.0$\times$ (tie) \\
          & Avg post ErrBits & \textbf{8.48} & 10.48 & \textbf{1.2$\times$ lower} \\
          & Max post ErrBits & \textbf{11.78} & 14.29 & \textbf{1.2$\times$ lower} \\
\bottomrule
\end{tabular}
\vspace{1pt}
\begin{minipage}{0.96\textwidth}
\footnotesize
\emph{Notes:} Bold marks the better value in each comparison, with ties bolded
on both systems. Post-error columns are deterministic 10,000-point sampled
repair-domain ErrBits averages and maxima.
\end{minipage}
\end{table}

\paragraph{Accuracy and refinement cost.}
The original failures are substantial: the median sampled peak inside the
repair domains is \textbf{52.00} ErrBits and the maximum is 62.94. Both systems
complete all \textbf{114/114} shared one-input configurations. With 10,000
samples per repair domain, \tool{}'s median post-repair average ErrBits are
1.00, 1.70, and 8.48 for \Heps{}, \Meps{}, and \Leps{}; the corresponding
sampled maxima are 4.16, 4.52, and 11.78. \autornp{} also finishes every case,
with median average/max post ErrBits of 4.78/7.18, 9.96/12.42, and 10.48/14.29.

Tightening the threshold changes refinement cost rather than success. From
\Leps{} to \Heps{}, \tool{}'s median repair time rises from \textbf{2.78}s to
\textbf{3.74}s, and one-piece coverage falls from 38/38 to 31/38 (37/38 at
\Meps{}); the strict tail reaches 69 pieces and 139.23s. \autornp{} also
repairs all cases, but its median repair time rises from 4.61s at \Leps{} to
81.97s at \Heps{}, and its median strict patch grows to 177.4KB. Across all
114 paired configurations, the \autornp{}/\tool{} repair-time ratio is
\textbf{3.33}$\times$ by geometric mean. Thus Taylor synthesis pays off most
when the requested accuracy forces the baseline to refine.

The strict \Heps{} setting shows the largest separation. On
\texttt{gsl\_sf\_airy\_Ai\_deriv\_scaled}, \autornp{} spends 6427.87s while
\tool{} finishes in 40.92s, a \textbf{157.1}$\times$ paired repair-time
speedup. The hard \texttt{gsl\_sf\_bessel\_Y1} row shows the same pattern:
9788.39s for \autornp{} versus 139.23s for \tool{}, or
\textbf{70.3}$\times$ faster synthesis. These are not additional successes:
both systems repair the rows. They show that the strict threshold turns
\autornp{}'s subdivision strategy into the bottleneck, while \tool{} reaches an
accepted local Taylor patch much earlier.

\paragraph{A representation boundary: the pole case.}
Both systems' ordinary polynomial backends fail for
\texttt{gsl\_sf\_psi\_1} around the pole at $x=-8$. With $t=x+8$, the
dominant $1/t^2$ term makes subdivision attack a singularity rather than a
smooth residual. Factoring that term and fitting
$\psi_1(x)\approx(1+t^2P(t))/t^2$ changes the representation, after which a
degree-20 repair passes all three thresholds with maximum checked error 1 bit
over 94,443--97,463 validation inputs. The singular point itself remains with
the original routine. This diagnostic is excluded from the standard aggregate:
it shows that failure is due to model mismatch, not merely insufficient search.

\paragraph{Multi-input frontier.}
PTB localizes \textbf{23/23} records at every threshold, so detection is not the
bottleneck. Synthesis repairs all \textbf{17/17} two-input cases at \Meps{} and
\Leps{} and 15/17 at \Heps{}, but only one of two three-input cases at the two
looser thresholds and no four-input cases. The sharp drop after localization
matters: it attributes the limit to covering the resulting box under a fixed
piece budget, rather than to an inability to find the high-error region.

\begin{table}[H]
\centering
\caption{Multi-input localization, repair cost, patch size, and runtime impact.
Times and patch sizes are medians over successful repairs; runtime columns report
median original/repaired speedup.}
\label{tab:multi-input-direct-summary}
\scriptsize
\setlength{\tabcolsep}{2.7pt}
\begin{tabular}{@{}lrrrrrrrrrr@{}}
\toprule
& \multicolumn{5}{c}{Coverage} & \multicolumn{2}{c}{Repair artifact} &
\multicolumn{3}{c}{Runtime speedup} \\
\cmidrule(lr){2-6}\cmidrule(lr){7-8}\cmidrule(l){9-11}
\metric{Threshold} & \metric{Loc.} & \metric{2D} & \metric{3D} & \metric{4D} &
\metric{All} & \metric{Time (s)} & \metric{Size (KB)} & \metric{Inside} &
\metric{Mixed} & \metric{Outside} \\
\midrule
\Heps{} & 23/23 & 15/17 & 0/2 & 0/4 & 15/23 & 12.73 & 22.4 & 1.92$\times$ & 1.33$\times$ & 0.99$\times$ \\
\Meps{} & 23/23 & 17/17 & 1/2 & 0/4 & 18/23 & 14.78 & 22.4 & 2.21$\times$ & 1.38$\times$ & 0.99$\times$ \\
\Leps{} & 23/23 & 17/17 & 1/2 & 0/4 & 18/23 & 20.10 & 22.4 & 2.20$\times$ & 1.38$\times$ & 0.99$\times$ \\
\bottomrule
\end{tabular}
\end{table}

\Cref{tab:multi-input-direct-summary} separates synthesis cost from execution
cost. Successful direct repairs take 12.73--20.10s at the median and emit
22.4KB median C patches at every threshold. Once installed, these patches speed
up inputs inside the repair box by 1.92--2.21$\times$ and mixed streams by
1.33--1.38$\times$, while inputs outside the box remain at 0.99$\times$ because
they only pay the guard before returning to the original routine.

\rqanswer{RQ1 answer}{The experiments demonstrate that \tool{} effectively
repairs localized high-error cases on the \gsl{} benchmark workload: it
completes all \textbf{114/114} shared one-input configurations with lower
median post-repair error than \autornp{}.
The advantage is strongest at the strict \Heps{} threshold, where \tool{} is
\textbf{21.9}$\times$ faster and emits \textbf{84.5}$\times$ smaller patches;
the multi-input results further show that \tool{} supports localized multi-D
benchmark cases.}

\subsection{RQ2: Repair Quality and Runtime Impact}
\label{sec:evaluation-rq2}

RQ2 asks whether a successful repair remains practical as code. The central
result is a separation of concerns: stricter or larger domains can make
\emph{synthesis} expensive, but accepted patches stay compact in the common
case and have a stable runtime profile.

\paragraph{Patch size follows refinement.}
The one-piece counts (31/38, 37/38, and 38/38 from strict to loose) explain the
size distribution better than a single overall median. At \Heps{}, a few hard
regions expand to as many as 69 pieces; at \Meps{} the maximum is three, and at
\Leps{} every repair is one piece. The paired 38-function patches therefore
show the same crossover as repair time: \tool{} is much smaller at \Heps{} and
\Meps{}, but the advantage narrows when both methods emit small \Leps{} patches.
Across all 114 paired configurations, \tool{} uses \textbf{0.162}$\times$ the
C bytes of \autornp{} by geometric mean.

The worst strict patch illustrates the difference. At \Heps{},
\texttt{gsl\_sf\_lngamma} is the largest patch for both systems:
\autornp{} emits \textbf{5.90}MB of C code (37,102 generated lines), whereas
\tool{} emits \textbf{47.5}KB across 69 pieces, a \textbf{124.1}$\times$
smaller maximum patch. The next-largest strict patch,
\texttt{gsl\_sf\_airy\_Ai\_deriv\_scaled}, is similarly separated: 5.36MB for
\autornp{} versus 20.3KB for \tool{}.

\paragraph{Runtime cost is localized by the guard.}
The full 38-function dual-backend runtime matrix has 684 successful rows:
38 functions, three thresholds, two systems, and inside/outside/mixed streams.
\tool{}'s geometric-mean speedups are \textbf{2.90}$\times$ inside the repair
domain, \textbf{1.34}$\times$ on mixed streams, \textbf{0.98}$\times$ outside,
and \textbf{1.56}$\times$ overall. \autornp{} reaches
\textbf{2.21}$\times$, \textbf{1.14}$\times$, \textbf{0.97}$\times$, and
\textbf{1.35}$\times$, respectively. Thus \tool{}'s repaired code has a
\textbf{1.16}$\times$ overall geometric-mean runtime advantage on the matched
workload, while both systems remain near parity outside the guard.

At \Heps{}, the maximum single-row execution speedup for \tool{} is
\textbf{11.34}$\times$ inside the guard on the scaled Airy-Ai row. \autornp{}
has one larger isolated inside speedup, \textbf{12.87}$\times$ on Airy-Ai, but
its strict aggregate is weaker: the \Heps{} overall geometric mean is
\textbf{1.39}$\times$ for \tool{} and \textbf{0.94}$\times$ for \autornp{},
with mixed-stream geometric means of \textbf{1.23}$\times$ and
\textbf{0.81}$\times$. Thus the strict-threshold advantage is primarily a
repair-cost and patch-size advantage, with a positive aggregate runtime profile
rather than a claim that every individual runtime row is faster.

\paragraph{Larger boxes can reduce synthesis success.}
\Cref{tab:multi-input-cap-sensitivity} varies the tuple-count cap on a 20-row
ordinary subset, excluding the known long-tail \texttt{gsl\_sf\_poch},
\texttt{gsl\_sf\_pochrel}, and \texttt{gsl\_sf\_hyperg\_U} cases. Increasing
the cap admits a larger repair box; it does not relax the error target or add a
larger piece budget.

\begin{table}[H]
\centering
\caption{Multi-input sensitivity to the repair-domain tuple-count cap. Median
times are \Heps{}/\Meps{}/\Leps{} in seconds; LT is a manually interrupted
long tail. A larger box can exhaust the fixed synthesis budget.}
\label{tab:multi-input-cap-sensitivity}
\scriptsize
\setlength{\tabcolsep}{3.2pt}
\begin{tabularx}{\textwidth}{@{}lcccc>{\raggedright\arraybackslash}X@{}}
\toprule
\metric{Cap} & \Heps{} & \Meps{} & \Leps{} & \metric{Med. time (s)} & Outcome \\
\midrule
$10^{10}$ & 20/20 & 20/20 & 20/20 & 28.6 / 25.9 / 25.2 & All repaired; max 1 piece \\
$10^{11}$ & 18/20 & 20/20 & 20/20 & 49.8 / 29.5 / 29.6 & Strict \texttt{H3d} failures; max 999 pieces \\
$10^{12}$ & 17/20 (+1 LT) & 19/20 & 20/20 & 46.4 / 46.3 / 25.5 & Strict \texttt{H3d} failures and one 4D hypergeometric long tail; max 999 pieces \\
\bottomrule
\end{tabularx}
\end{table}

The counterintuitive result is monotone pressure in the wrong direction: the
$10^{10}$ cap repairs all 60 threshold configurations with one piece, whereas
$10^{11}$ loses two strict \texttt{H3d} cases at 999 pieces and $10^{12}$ also
introduces a four-input long tail. Larger boxes expose more tuples and local
variation while the 1000-piece budget stays fixed. Together with the direct
multi-input runtime data in \cref{tab:multi-input-direct-summary}, this
identifies the cap as a synthesis tractability control rather than an accuracy
or execution-performance knob.

\rqanswer{RQ2 answer}{The accepted repairs are high quality as generated code:
\tool{} patches are compact, faster inside the repair guard, and near parity
outside it. Strict thresholds and larger multi-input boxes mainly stress
synthesis, but once found, the patches retain a favorable runtime profile.}

\subsection{Limitations and Threats to Validity}
\label{sec:evaluation-limitations}

The repair checker is sampled and local. It validates the emitted patch on the
recorded repair-domain samples, endpoints, and pre-repair peak input, but it
does not prove the full surrounding original routine or continuity across
adjacent pieces. The experiments also check the scalar return value, not all
possible library side channels such as status codes, \texttt{errno}, or
floating-point exception flags.

Taylor polynomials are not the right representation for every localized
failure. The \texttt{gsl\_sf\_psi\_1} pole case requires factoring the dominant
singular term before polynomial repair becomes effective. This points to a
natural extension: detect singular or rational structure and choose a local
representation before applying the Taylor search.

Finally, timing comparisons depend on compiler, platform, and input-stream
construction. We report matched workloads, shared PTB domains, repeated timed
calls, and geometric means, but the results should be read as evidence for the
evaluated environment rather than as platform-independent constants.

%% file: sections/06-related-work.tex
\section{Related Work}
\label{sec:related-work}

\subsection{Automated Floating-Point Repair and Accuracy Improvement}
\label{sec:related-repair}

Automated repair systems search for source-level or expression-level changes
that improve a program's behavior. General automated-program-repair techniques
synthesize patches from tests, constraints, angelic values, learned correctness
models, or search
\cite{weimer2009automatically,nguyen2013semfix,Mechtaev2015DirectFixLF,Mechtaev2016AngelixSM,xuan2017nopol,long2016automatic}.
For floating-point programs, prior work also finds inputs that expose high
error or instability
\cite{conf/icse/ZouWXZSM15,yi2017efficient,conf/ppopp/ChiangGRS14}.
FPCC detects high-error inputs using chain conditions~\cite{fpcc}; ATOMU uses
atomic conditions~\cite{zoudetecting}; FPGen uses symbolic execution to
generate error-inducing inputs~\cite{DBLP:conf/wcre/YuYYLCH23}; CoverMe
targets high coverage for floating-point code~\cite{Fu:2017:AHC:3062341.3062383};
and dynamic analyses localize or explain numerical instability and root causes
\cite{Bao:2013:ODI:2509136.2509526,benz2012dynamic,TangTSE16,sanchez2017finding}.
Herbie rewrites expressions to improve accuracy
\cite{panchekha2015automatically}, while transformation and tuning systems
change expression structure, precision, or mixed-precision assignments
\cite{ICSE/2019/wang,Martel:2009:PTN:1480945.1480960,damouche2017improving,damouche2017salsa,Precimonious,schkufza2014stochastic,lam2013automatically}.

\autornp{} is our matched repair baseline because it targets numerical-library
errors and supplies the same PTB-localized domains used in our comparison
\cite{yi2017automated,Yi:2019:EAR:3302515.3290369}. Methodologically,
ACESO and OFP-Repair are oracle-free repair tools that synthesize patches using
only the original precision~\cite{zou2022oraclefree,tan2025ofprepair}. However,
both exclude ill-conditioned problems: ACESO explicitly places them outside
its scope, while OFP-Repair rejects cases whose condition numbers exceed its
threshold. The high-error \gsl{} programs detected in our evaluation all exceed
this threshold. In contrast, \tool{} supports such ill-conditioned cases by
using high-precision coefficients together with EFT-based transformations.

\subsection{Polynomial Approximation and Library Implementation}
\label{sec:related-approx}

Taylor, Chebyshev, Remez, and minimax approximations are classical tools for
mathematical-function implementation
\cite{hart1968computer,Abramowitz:1974:HMF:1098650,Knuth:1997:ACP:270146,muller2016elementary}.
Range reduction, table-driven reconstruction, and special-case handling are
also standard ingredients in numerical libraries
\cite{tang1989table,payne1983radian,muller2016elementary}. Horner and
multivariate Horner schemes organize polynomial evaluation efficiently
\cite{cajori1911horner,mulhorner,greedyhorner}, and compensated arithmetic,
error-free transformations, and fused multiply-add can improve the behavior of
floating-point evaluation
\cite{Dekker:1971:FTE:2716631.2717032,accdotpro,fmaIEEE}.

\tool{} uses these standard ingredients rather than proposing a new
approximation theory. Its use of Taylor polynomials is local and operational: a
Taylor expression becomes an emitted floating-point C fragment with a fixed
evaluation order, an explicit guard, a fallback path, and a repair-specific
checker. The contribution is the checked repair loop around this representation;
the \gsl{} experiments instantiate this loop on localized library errors and
measure its repair cost and patch size against a matched baseline.

\subsection{Floating-Point Analysis and Verification}
\label{sec:related-verification}

Floating-point analysis tools, testing frameworks, and verification systems
provide techniques for finding high-error inputs, bounding roundoff error, or
checking numerical properties. Interval arithmetic, abstract interpretation,
and static numerical analysis provide foundations for bounding numerical
behavior~\cite{hickey2001interval,cousot1977abstract,putot2004static,goubault2006static,darulova2011trustworthy}.
FPTaylor uses symbolic Taylor expansions to estimate roundoff error and can
emit proof certificates~\cite{10.1145/3230733}. Gappa-style certification and
verified numerical pipelines address proof obligations for floating-point code
and approximations~\cite{deDinechin2011gappa,becker2022dandelion}. Other work
verifies or tunes floating-point and math-library code
\cite{lee2017automatically,chiang2017rigorous,conf/oopsla/FuBS15}.

\tool{} is complementary to this line of work. It is not a general-purpose
floating-point verifier and does not emit proof certificates like FPTaylor.
Instead, it generates concrete repair code and relies on the task-specific
checker stated in \cref{sec:problem}: local threshold validation over recorded
repair-domain inputs.

%% file: sections/08-conclusion.tex
\section{Conclusion}
\label{sec:conclusion}

This paper presented \tool{}, a Taylor-approximation-based framework for
repairing localized high floating-point errors in numerical C routines.
\tool{} uses localized high-error domains to synthesize guarded Taylor patches,
lowers each candidate to concrete C with fixed floating-point evaluation order,
and accepts a repair only after validating the emitted patch on the recorded
repair-domain inputs.

The evaluation shows that \tool{} and \autornp{} both repair all 114 one-input
\gsl{} benchmark configurations, but \tool{} does so with lower cost:
3.33$\times$ geometric-mean repair-time advantage, 0.162$\times$ as many C
bytes, and a 157.1$\times$ maximum paired repair-time speedup in the strictest
setting. The strict-level maximum patch also shrinks from 5.90MB with
\autornp{} to 47.5KB with \tool{}. The repaired-code runtime matrix is
consistent with this picture in aggregate, with 1.56$\times$ overall
geometric-mean speedup over the unpatched implementations for \tool{} and
1.35$\times$ for \autornp{}. \tool{} also repairs bounded multi-input benchmark
cases that are outside \autornp{}'s repair workflow. These results support localized
one-input repair and a bounded multi-input capability analysis; full-domain
proofs, arbitrary multi-input repair, and automatic singular-representation
selection remain outside the present claim.

%% file: sections/09-data-availability.tex
\section*{Data Availability}

An artifact accompanies this submission. It includes the \tool{}
source code, scripts for the \gsl{} benchmark experiments, Docker build files,
benchmark harnesses, cached result tables used in \Cref{sec:evaluation}, and
instructions for reproducing the main experiments. The artifact is organized so
that fresh reproduction runs write to separate output directories and do not
overwrite the bundled paper data. We will make the artifact publicly available
upon publication.